\documentclass{egpubl}
\usepackage{eurovis2019}

 \electronicVersion 

\ifpdf \usepackage[pdftex]{graphicx} \pdfcompresslevel=9
\else \usepackage[dvips]{graphicx} \fi

\PrintedOrElectronic

\usepackage{t1enc,dfadobe}

\usepackage{egweblnk}
\usepackage{cite}

\usepackage[usenames,dvipsnames]{xcolor}
\usepackage[normalem]{ulem}
\definecolor{mred}{rgb}{.80,.12,.30}
\definecolor{grey}{rgb}{0.5,0.5,0.5}
\definecolor{outlinecolor}{rgb}{0,0.18,0.38}
\definecolor{Purple}{rgb}{.75,0,.85}

\newif\ifnotes
\notestrue

\let\origcite\cite
\renewcommand{\cite}[1]{\ifnotes\mbox{\origcite{#1}}\else \origcite{#1}\fi}

\title[VA for EMA]%
      {A User-based Visual Analytics Workflow for Exploratory Model Analysis}

\author[D. Cashman, S. R. Humayoun, F. Heimerl, K. Park, S. Das, J. Thompson, B. Saket, A. Mosca, J. Stasko, A. Endert, M. Gleicher, R. Chang]
{\parbox{\textwidth}{\vspace{-3em}\centering Dylan Cashman$^{1,*}$\orcid{0000-0003-4853-5701},
        Shah Rukh Humayoun$^{1,*}$\orcid{0000-0002-4645-8223}, 
        Florian Heimerl$^2$\orcid{0000-0002-3943-2260}, Kendall Park$^2$\orcid{0000-0002-7302-0139}, Subhajit Das$^3$\orcid{0000-0002-4065-8793}, John Thompson$^3$\orcid{0000-0002-3102-4035}, Bahador Saket$^3$\orcid{0000-0002-5896-0149}, Abigail Mosca$^1$\orcid{0000-0002-9008-5516}, John Stasko$^3$, Alex Endert$^3$\orcid{0000-0002-6914-610X}, Michael Gleicher$^2$\orcid{0000-0003-3295-4071}, and Remco Chang$^1$\orcid{0000-0002-6484-6430}
        }
        \\
{\parbox{\textwidth}{\vspace{-1em}\centering $^1$Tufts University, USA \hspace{4mm}
         $^2$Georgia Tech, USA \hspace{4mm}
         $^3$University of Wisconsin -- Madison, USA \hspace{4mm}
         $^*$These two authors contributed equally
       }
}
}

\begin{document}

\teaser{
    \vspace{-3em}
  \centering
  \includegraphics[width=0.95\linewidth]{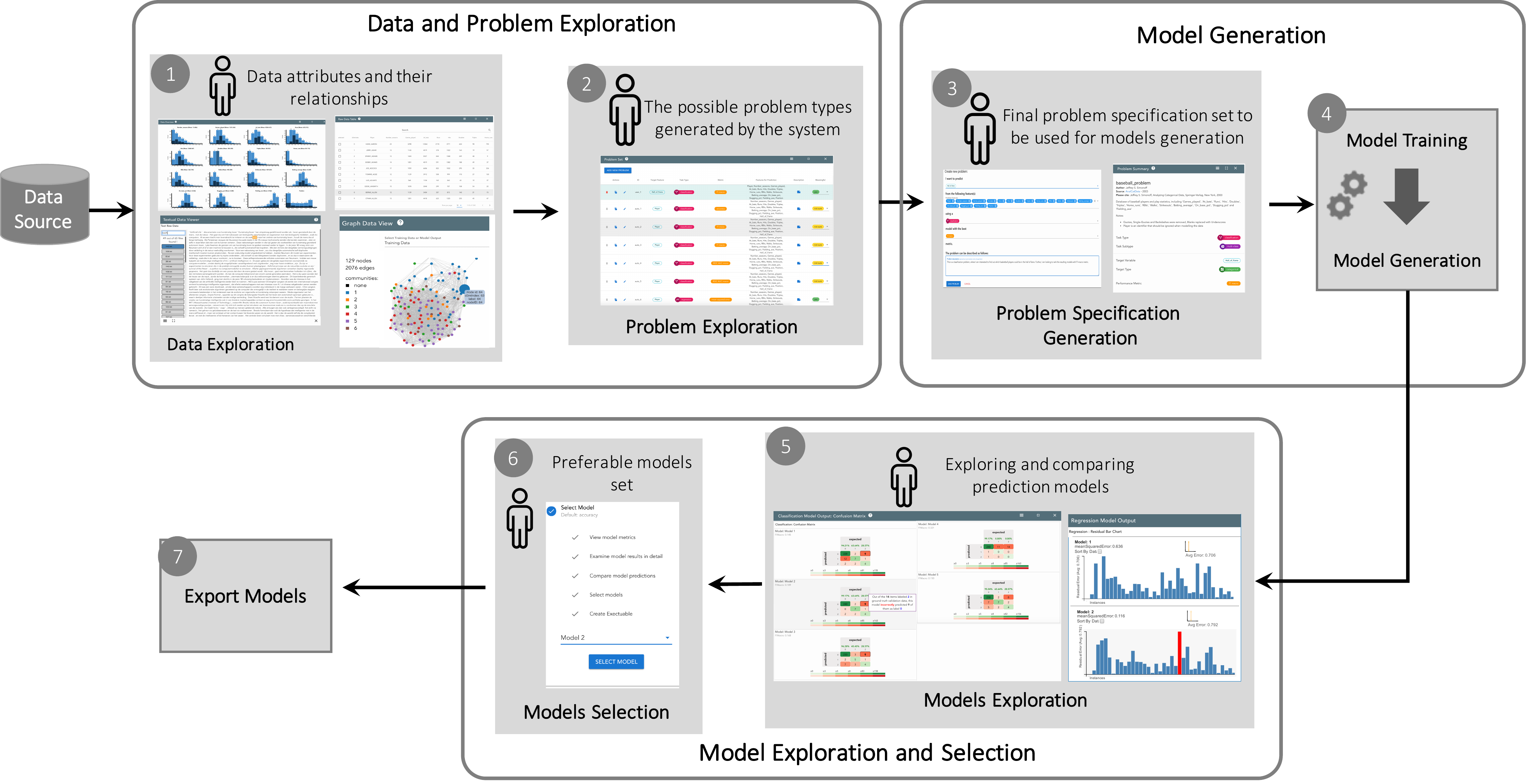}
  \caption{The proposed EMA visual analytics workflow for discovery and generation of machine learning models. In \textbf{step 1}, the system uses interactive visualizations (such as histograms or graphs) to provide an initial data overview. The system then generates a number of possible modeling problems based on analyzing the data set (\textbf{step 2}) from which the user analyzes and selects one to try (\textbf{step 3}).  Next, (\textbf{step 4}) an automated ML system trains and generates candidate models based on the data set and given problem. In \textbf{step 5}, the system shows comparisons of the  generated prediction models through interactive visualizations of their predictions on a holdout set. Lastly, in \textbf{step 6}, users can select a number of preferable models, which are then exported by the system during \textbf{step 7} for predictions on unseen test data.  At any time, users can return to \textbf{step 3} and try different modeling problems on the same dataset.}
	\label{fig:workflow}
}

\maketitle
\begin{abstract}



Many visual analytics systems allow users to interact with machine learning models towards the goals of data exploration and insight generation on a given dataset. However, in some situations, insights may be less important than the production of an accurate predictive model for future use.  In that case, users are more interested in generating of diverse and robust predictive models, verifying their performance on holdout data, and selecting the most suitable model for their usage scenario.
In this paper, we consider the concept of Exploratory Model Analysis (EMA), which is defined as the process of discovering and selecting relevant models that can be used to make predictions on a data source.  We delineate the differences between EMA and the well-known term exploratory data analysis in terms of the desired outcome of the analytic process: insights into the data or a set of deployable models. The contributions of this work are a visual analytics system workflow for EMA, a user study, and two use cases validating the effectiveness of the workflow.  We found that our system workflow enabled users to generate complex models, to assess them for various qualities, and to select the most relevant model for their task. 

\end{abstract}  
\section{Introduction}

Exploratory data analysis (EDA) has long been recognized as one of the main components of visual analytics \cite{cook2005illuminating}.
EDA is an analysis process through which a user ``\textit{searches and analyzes databases to find implicit but potentially useful information}''~\cite{keim2006challenges}, with the use of an interactive visual interface.
As described by Tukey, the process of data exploration helps users to escape narrowly assumed properties about their data and allows them to discover patterns and characteristics that were not previously known~\cite{tukey1977exploratory}.
In this sense, the goal of EDA and the use of traditional visual analytics systems is to help the user gain early insight into their data \cite{north2006toward, chang2009defining}.

However, in the modern era of big data, machine learning, and AI, visual analytics systems have begun to take on a new role: to help the user in refining \textit{machine learning models}. Systems such as TreePOD\cite{muhlbacher2018treepod}, BEAMES\cite{dasbeames}, and Seq2SeqVis\cite{strobelt2018seq2seq} propose new visualization and interaction techniques not for a user to better understand their data, but to understand the characteristics of the machine learning models trained on their data and the effects of modifying their parameters and hyperparameters.
The goal of these visual analytics systems is to produce a predictive model which will then be used on unseen data.

These systems help analyze and refine a particular type of model with a predefined modeling goal.  This limits their ability to support an exploratory analysis process since the user cannot try multiple modeling problems in the same system, and instead are confined to decision trees, regressions, and sequence-to-sequence models, respectively.
In this work, we consider a previously unsupported scenario in which the \textit{type of model and the modeling task is not known} at the beginning of the analysis.
We introduce the term \textit{Exploratory Model Analysis} (EMA), and define it as the process of exploring the set of potential models that can be trained on a given set of data.
EMA shares characteristics with EDA in that both describe an analysis process that is open-ended and whose results are not clearly defined a priori, and may change and adapt during the process.

The goal of EMA is twofold: discover variables in the dataset on which reliable predictions can be made, and find the most suitable and robust types of models to predict these variables.
There may be multiple models discovered at the end of the process - an analyst may end up discovering regression models between variables $a$, $b$, and $c$, classification models where variables $d$ and $e$ predict the label of variable $f$, and neural networks that use all independent variables to predict the value of variable $g$.  

Despite the parallels between the two, the analysis processes that EDA and EMA describe are applicable to different sets of analysis scenarios.
To illustrate the difference, consider two users of visual analytics systems in a financial services company: a broker, who must be able to explain the current state of the market, in the context of its near present and past, and the quantitative analyst, who must be able to model the future behavior of the market. The broker may use machine learning models to support their exploration of the data, but their ultimate goal is to understand current patterns in the data, so that they can make decisions in the current market landscape.  
In contrast, the quantitative analyst might be interested in what types of predictions are possible given the data being collected, and beyond that, which types of predictions are robust.  Exploratory data analysis might expose some information that is predictive, such as the correlation between features, but for large and complex datasets, complex modeling is needed to make sufficiently robust predictions.  The use of our visual analytics workflow can help the quantitative analyst to try different types of models and explore the model space.  


In this example, there are two distinctions between these two users: (1) their intended goals, and (2) how data is used in the process. For the broker, the intended outcome of using visual analytics is a decision, a data item (e.g. in an anomaly detection task), or an interesting pattern within the data. The data is therefore the focus of the investigation. On the other hand, for an analyst, the intended outcome is a model (or set of models), its hyperparameters, and properties about its predictions on held out data.  The data is used to train and validate the model. It is not in itself the focus of attention.


While there is a plethora of tools and techniques in the visual analytics literature that support using machine learning models, most existing workflows (such as the visual data-exploration workflow by Keim et al.\cite{Keim2008}, the knowledge generation model by Sacha et al.\cite{sacha2014knowledge}, the economic model of visualization by van Wijk\cite{van2005value}, and four out of the six workflows described by Chen and Golan~\cite{Chen2016}) focus on the exploration and analysis of data, rather than the discovery of the model itself.  These workflows presuppose that the user knows what their modeling goal was (e.g. using a regression model to predict the number of hours a patient will use a hospital bed).
Although these workflows (and the many visual analytics systems built following these workflows) are effective in helping a user in data exploration tasks, we note that there is often an earlier step of modeling where users do not yet know what types of models can be built from a data source.
Model exploration is an important aspect of data analysis that is underrepresented in visual analytics workflows.  We do note that the two \textit{Model-developmental Visualization} workflows from Chen and Golan do consider the goal of exporting a model rather than analyzing data\cite{Chen2016}.  However, they are not described in detail and only provided as abstractions.  In contrast, this work delves deeply into each step of its workflow, and provides an example of its implementation.

The primary contribution of this work is a workflow for EMA that supports model exploration and selection.
We first identified a set of functionality and design requirements needed for EMA through a pilot user study.
These requirements are then synthesized into a step-by-step workflow (see Figure~\ref{fig:workflow}) that can be used to implement a system supporting EMA. To validate our proposed workflow for exploratory model analysis, we developed a prototype visual analytics system for EMA and ran a user study with nine data modelers. We report the outcomes of this study and also present two use cases of EMA to demonstrate its applicability and utility.

To summarize, in this paper we make contributions to the visual analytics community in the following ways:

\begin{itemize}
\item \textbf{Definition of exploratory model analysis}: We introduce the notion of exploratory model analysis and propose an initial definition.
\item \textbf{Workflow for exploratory model analysis}: Based on a pilot study with users, we developed a workflow that supports exploratory model analysis.
\item \textbf{User studies that validate the efficacy and feasibility of the workflow}: We developed a prototype visual analytics system based on our proposed workflow and evaluated its efficacy with domain expert users.  We also present two use cases to illustrate the use of the system.
\end{itemize}

\section{Related Work}


\subsection{Exploratory Data Analysis}
The statistician Tukey developed the term \textit{exploratory data analysis} (EDA) in his work from 1971 through 1977\cite{tukey1993exploratory} and his 1977 book of the same name\cite{tukey1977exploratory}.  EDA focuses on exploring the underlying data to isolate features and patterns within\cite{Hoaglin1983}.  EDA was considered a departure from standard statistics in that it de-emphasized the methodical approach of posing hypotheses, testing them, and establishing confidence intervals\cite{church1979look}.  Tukey's approach tended to favor simple, interpretable conclusions that were frequently presented through visualizations.  


A flourishing body of research grew out of the notion that visualization was a critical aspect of making and communicating discoveries during EDA~\cite{perer2008integrating}. This includes (static) statistical visualization libraries (such as ggplot\cite{wickham2008ggplot2}, plotly\cite{sievert2016plotly}, and matplotlib\cite{hunter2007matplotlib}), visualization libraries (such as D3\cite{bostock2011d3}, Voyager\cite{satyanarayan2017vega}, InfoVis toolkit\cite{fekete2004infovis}), commercial visualization systems (such as Tableau\cite{tableau}, spotfire\cite{spotfire}, Power BI\cite{powerbi}), and other visualization software designed for specific types of data or domain applications (for some examples, see surveys such as\cite{de2003visual,heer2010tour}).

\subsection{Visual Analytics Workflows}
Visual analytics workflows\footnote{Frameworks, pipelines, models, and workflows are often used interchangeably in the visualization community to describe abstractions of sequences of task. In this paper, we use the word \textit{workflow} to avoid confusion. Further, we use the word \textit{model} to specifically refer to machine learning models and not visualization workflows.}, including the use of models, grew out of research into Information Visualization (Infovis).
Chi and Riedl\cite{chi1998operator} proposed the \textit{InfoVis reference model} (later refined by Card, Mackinlay and Shneiderman\cite{card1999readings}) that emphasizes the mapping of data elements to visual forms.
The framework by van Wijk\cite{van2005value} extends this with interaction -- a user can change the specification of the visualization to focus on a different aspect of the data.

The notion of effective design in Infovis has largely been summarized by Shneiderman's mantra; \textit{Overview, zoom \& filter, details-on-demand}~\cite{shneiderman1996eyes}. Keim et. al. noted that as data increases in size and complexity, it becomes difficult to follow such a mantra; an overview of a large dataset necessitates some sort of reduction of the data by sampling or, alternatively, an analytical model. The authors provide a framework of Visual Analytics that incorporates analytical models in the visualization pipeline~\cite{Keim2010}. Wang et al.~\cite{Wang2016} extended the \textit{models} phase in the framework by Keim et al. to include a model-building process  with: \emph{feature selection and generation}, \emph{model building and selection}, and \emph{model validation}.
Chen and Golan~\cite{Chen2016} discuss prototypical workflows that include model building to aid data exploration with various degrees of model integration into the analysis workflows.
Sacha et al. formalized the notion of user knowledge generation in visual analytics system, accounting for modeling in the feedback loop of a mixed-initiative system~\cite{Sacha16}. While these frameworks have proven invaluable in guiding the design of countless visual analytics systems, they muddle the delineation between the different goals of including modeling in the visualization process, conflating model building with insight discovery.  

The ontology for visual analytics assisted machine learning proposed by Sacha et al.~\cite{Sacha2019} offers the clearest background on which to describe our workflow's application to EMA.  In that work, the authors present a fairly complete knowledge encoding of common concepts in visual analytics systems that use machine learning, and offer suggestions of how popular systems in the literature map onto that encoding.  While each step of our workflow can be mapped into the ontology, a key distinction in our workflow is in the \textit{Prepare-Learning} process.  The authors note that "in practice, quite often, the ML \textit{Framework} was determined before the step \textit{Prepare-Data} or even before the raw data was captured", and, in fact, none of the four example systems in that work explicitly use visual analytics to support the step of choosing a model.  In our workflow, this is not the case - the framework, or machine learning modelling problem and its corresponding algorithms are not chosen a priori.  We also note that the term EMA itself could comprise the entirety of the VIS4ML ontology, as each step could be useful in exploratory modeling.  In that case, our workflow does not completely support EMA; such a system would need to support every single step of the ontology.  However, in our definition, the choice of modeling problem is an necessary condition for EMA, and thus, our workflow is the first that is sufficient for supporting EMA with visual analytics.

\begin{figure}[t]
\begin{center}
\centering
 \includegraphics[width=1\linewidth]{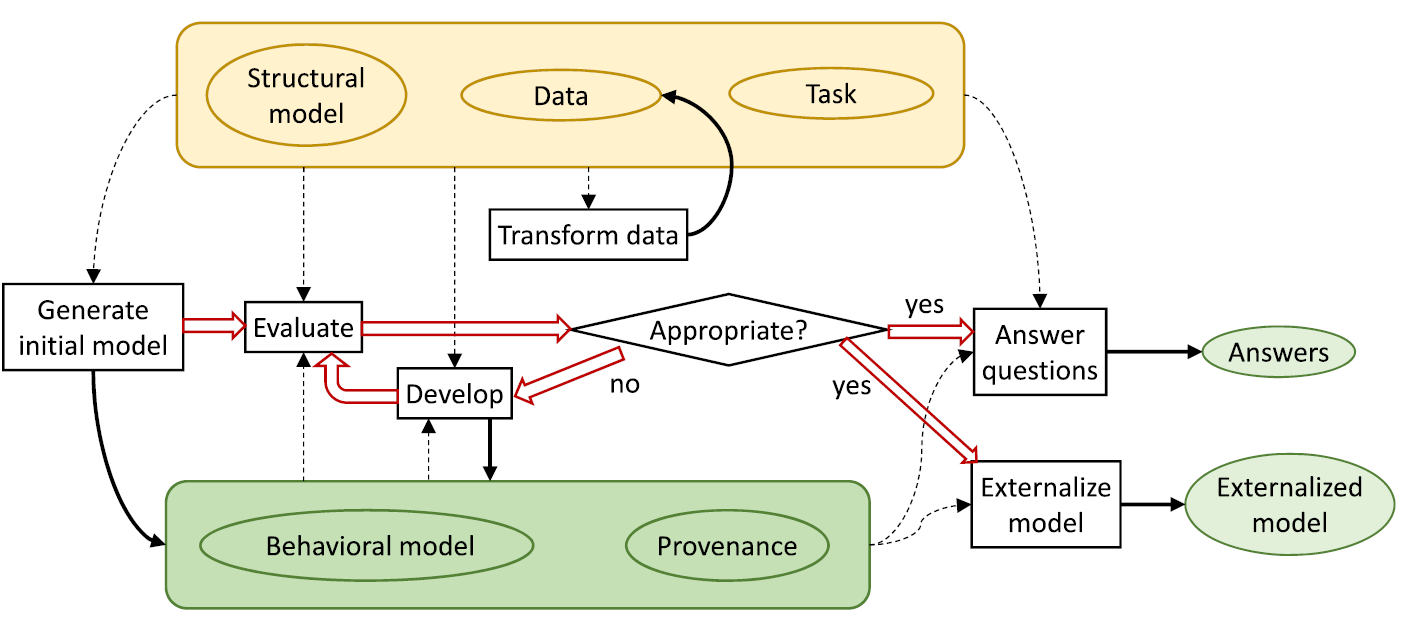}
 \caption{The model generation framework of visual analytics by Andrienko et al.\cite{Andrienko2018}.}
 \label{fig:andrinko}
\end{center}
\end{figure}

Most similar to our proposed EMA workflow is the one recently introduced by Andrienko et al.\cite{Andrienko2018} that posits that the outcome of the VA process can either be an ``answer'' (to a user's analysis question) or an ``externalized model''.  Externalized models can be deployed for a multitude of reasons, including automating an analysis process at scale or for usage in recommender systems. While similar in concept, we propose that the spirit of the workflow by Andrienko et al. is still focused on data exploration (via model generation) which does not adequately distinguish between a data- from a model-focused use case such as the aforementioned financial broker and the quantitative analyst. In a way, our EMA workflow can be considered as the process that results in an \textit{initial model}, which can then be used as input to the model by Andrienko et al. (i.e. box (7) in Figure \ref{fig:workflow} as the input to the first box in the Andrienko model shown in Figure \ref{fig:andrinko}). 

\subsection{Modeling in Visual Analytics}
We summarize several types of support for externalizing models using visual analytics with a similar categorization to that given by Liu et. al.~\cite{liu2017towardsmlVA}.  We summarize these efforts into four groups: visual analytics for model \textbf{explanation}, \textbf{debugging}, \textbf{construction and steering}, and \textbf{comparison}, noting how they differ from our definition of EMA.

\noindent \textbf{Model Construction and Steering.}
A modeling expert frequently tries many different settings when building a model, modifying various hyperparameters in order to maximize some utility function, whether explicitly or implicitly defined. Visual analytics systems can assist domain experts to control the model fitting process by allowing the user to directly manipulate the model's hyperparameters or by inferring the model's hyperparameters through observing and analyzing the user's interactions with the visualization.

Sedlmair et al.\cite{sedlmair2014visual} provide a comprehensive survey of visual analytics tools for analyzing the parameter space of models.
Example types of models used by these visual analytics tools include regression\cite{muhlbacher2013partition}, clustering\cite{nam2007clustersculptor, cavallo2018clustrophile, kwon2018clustervision, Sacha2018}, classification\cite{van2011baobabview, choo2010ivisclassifier}, dimension reduction\cite{choo2013interactive, jeong2009ipca, nam2013tripadvisor, anand2012visual, liu2015visual}, and domain-specific modeling approaches including climate models\cite{Wang2017}.
In these examples, the user directly constructs or modifies the parameters of the model through the interaction of sliders or interactive visual elements within the visualization.

In contrast, other systems support model steering by inferring a user's interactions. Sometimes referred to as semantic interaction\cite{endert2012semantic}, these systems allow the user to perform simple, semantically relevant interactions such as clicking and dragging and dynamically adjusts the parameters of the model accordingly.
For example, 
ManiMatrix is an interactive system that allows users to express their preference for where to allot error in a classification task\cite{kapoor2010interactive}. By specifying which parts of the confusion matrix they don't want error to appear in, they tell the system to search for a classification model that fits their preferences. Disfunction\cite{brown2012dis} allows the user to quickly define a distance metric on a dataset by clicking and dragging data points together or apart to represent similarity. Wekinator enables a user to implicitly specify and steer models for music performance\cite{fiebrink2009meta}. BEAMES\cite{dasbeames} allows a user to steer multiple models simultaneously by expressing priorities on individual data instances or data features. Heimerl et. al.\cite{heimerl2012visual} support the task of refining binary classifiers for document retrieval by letting users interactively modify the classifier's decision on any document. 

\noindent \textbf{Model Explanation.} The explainability of a model is not only important to the model builder themselves, but to anyone else affected by that model, as required by ethical and legal guidelines such as the European Union's General Data Protection Regulation (GDPR)\cite{eu-269-2014}. Systems have been built to provide insight into how a machine learning model makes predictions by highlighting individual data instances that the model predicts poorly.  With Squares\cite{ren2017squares}, analysts can view classification models based on an in-depth analysis of label distribution on a test data set.  Krause et. al. allowed for instance-level explanations of triage models based on a patient's medication listed during intake in a hospital emergency room\cite{krause2017workflow}.  Gleicher noted that a simplified class of models could be used in a VA application to trade off some performance in exchange for a more explainable analysis\cite{gleicher2013explainers}. Many other systems and techniques purport to render various types of models interpretable, including deep learning models \cite{liu2018analyzing, liu2017towards, strobelt2018lstmvis, DBLP:journals/corr/YosinskiCNFL15, Bilal2018}, topic models\cite{wei2010tiara}, word embeddings \cite{heimerl2018interactive}, regression models\cite{muhlbacher2013partition}, classification models \cite{patel2010gestalt, ribeiro2016should,amershi2015modeltracker}, and composite models for classification~\cite{Liu2018boosting}.
While model explanation can be very useful in EMA, it does not help a user discover models, it only helps interpret them.  It is, however, a key tool in the exploration and selection of models (steps 5 and 6 of our workflow in Figure~\ref{fig:workflow}.

\noindent \textbf{Model Debugging.} While the calculations used in machine learning models can be excessively complicated, endemic properties of models that cause poor predictions can sometimes be diagnosed visually relatively easily.
RNNBow is a tool that uses intermediate training data to visually reveal issues with gradient flow during the training process of a recurrent neural network\cite{cashman2017rnnbow}.
Seq2Seq-Vis visualizes the five different modules used in sequence-to-sequence neural networks, and provides examples of how errors in all five modules can be diagnosed\cite{strobelt2018seq2seq}.
Alsallakh et al. provide several visual analysis tools for debugging classification errors by visualizing the class probability distributions\cite{alsallakh2014visual}.
Kumpf et al.~\cite{Kumpf2018} provide an interactive analysis method to debug and analyze weather forecast models based on their confidence estimates.
These tools allow a model builder to view \textit{how} and \textit{where} their model is breaking, on specified data instances.
Similar to model explanation, it incrementally improves a single model rather than discovers new models.

\noindent \textbf{Model Comparison.}
The choice of which model to use from a set of candidate models is highly dependent on the needs of the user and the deployment scenario of a model.  Gleicher provides strategies for accommodating comparison with visualization, many of which could be used to compare model outputs\cite{gleicher2018considerations}. Interactivity can be helpful in comparing multiple models and their predictions on a holdout set of data. Zhang et. al. recently developed Manifold, a framework for interpreting machine learning models that allowed for pairwise comparisons of various models on the same validation data\cite{zhang2018manifold}. M\"uhlbacher and Piringer\cite{muhlbacher2013partition} support analyzing and comparing regression models based on visualization of feature dependencies and model residuals.
TreePOD\cite{muhlbacher2018treepod} helps users balance potentially conflicting objectives such as accuracy and interpretability of decision tree models by facilitating comparison of candidate tree models.  Model comparison tools support model selection, but they assume that the problem specification that is solved by those models is already determined, and thus they do not allow exploration of the model space.

\section{A Workflow for Exploratory Model Analysis}

The four types of modeling described above all presuppose that the user's modeling task is well-defined: the user of the system already knows what their goal is in using a model. We contend that our workflow solves a problem that is underserved by previous research - Exploratory Model Analysis (EMA).  In EMA, the user seeks to discover what modeling can be done on a data source, and hopes to export models that excel at the discovered modeling tasks.  Some of the cited works do have some exploratory aspects, including allowing the user to specify which feature in the dataset is the target feature for the resulting predictive model.  However, to the best of our knowledge, no existing system allows for multiple modeling types, such as regression and classification, within the same tool.

Beyond the types of modeling outlined above, there are two new requirements that must be accounted for. First, EMA requires an interface for modeling problem specification - the user must be able to explore data and come up with relevant and valid modeling problems.  Second, since the type of modeling is not known \textit{a priori}, a common workflow must be distilled from all supported modeling tasks. All of the works cited above are specifically designed towards a certain kind of model, and take advantage of qualities about that model type (i.e. visualizing pruning for decision trees).  To support EMA, an application must support model discovery and selection in a general way.

In this section, we describe our method for developing a workflow for EMA. 
We adopt a user-centric approach that first gathers task requirements for EMA following similar design methodologies by Lloyd and Dykes\cite{lloyd2011human} and Brehmer et al.\cite{brehmer2014overview}.
Specifically, this design methodology calls for first developing a prototype system based on best practices.
Feedback by expert users are then gathered and distilled into a set of design or task requirements.
The expert users in this feedback study were identified by the National Institute of Standards and Technology (NIST) and were trained in data analysis. 
Due to confidentiality reasons, we do not report the identities of these individuals. 


\subsection{Prototype System}
\begin{figure}[htb]
\begin{center}
\centering
 \includegraphics[width=0.95\linewidth]{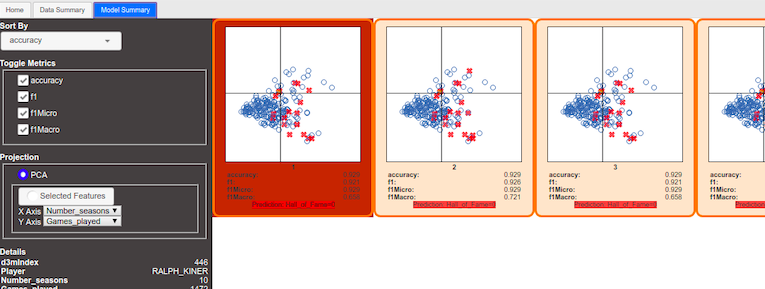}
 \caption{A prototype EMA visual analytics system used to determine task requirements. Classification is shown in this figure. During a feedback study with expert users, participants were asked to complete model selection tasks using this view. This process is repeated for regression models (not shown).  Feedback from this prototype was used to distill common steps in model exploration and selection across different model types.}
 \label{tab:pilotsystem}
\end{center}
\end{figure}

Our goal in this initial feedback study was to distill a common workflow between two different kinds of modeling tasks.  Our initial prototype system for supporting exploratory model analysis allowed for only two types of models -- classification and regression.
The design of this web-based system consisted of two pages using tabs, where on the first page, a user sees the data overview summary through an interactive histogram view. Each histogram in this view represented an attribute/field in the data set, where the x-axis represented the range of values while the y-axis represented the number of items in each range of values. On the second tab of the application, the system showed a number of resulting predicted models based on the underlying data set.  A screenshot of the second tab of this prototype system is shown in Figure~\ref{tab:pilotsystem}.  

Classification models were shown using scatter plots, where each scatter plot showed the model's performance on held out data, projected down to two dimensions. Regression models were visualized using bar charts, where each vertical bar represented the amount of residual and the shape of all the bars represents the model's distribution of error over the held out data. 

\subsection{Task Requirements}

We conducted a feedback study with four participants to gather information on how users discover and select models.  
The goal of the study was to distill down commonalities between two problem types, classification and regression,  
in which the task was to export the best predictive model.  
Each of the four participants used the prototype system to examine two datasets, one for a classification task and the other for regression.  Participants were tasked with exporting the best possible predictive model in each case.  
The participants were instructed to ask questions during the pilot study.  Questions as well as think-aloud was recorded for further analysis.
After each participant completed their task, they were asked a set of seven open-ended questions relating to the system's workflow, including what system features they might use for more exploratory modeling. 
The participants' responses were analyzed after the study and distilled into a set of six requirements for exploratory model analysis:

\begin{itemize}
    \item \emph{G1: Use the data summary to generate prediction models:} Exploration of the dataset was useful for the participants to understand the underlying dataset. This understanding can then be transformed into a well-defined problem specification that can be used to generate the resulting prediction models.  Visualization can be useful in providing easy exploration of the data, and cross-linking between different views into the dataset can help facillitate understanding and generate hypotheses about the data.
    \item \emph{G2: Change and adjust the problem specification 
    to get better prediction models:} Participants were interested in modifying the problem specifications to change the options (e.g., performance metrics such as accuracy, f1-macro, etc. or the target fields) so that they would get more relevant models.  The insights generated by visual data exploration can drive the user's refinements of the problem specification.
    \item \emph{G3: Initially rank the resulting prediction models:} Participants were interested to see the ranking of prediction models based on some criteria, e.g., a performance metric. The ranking should be meaningful to the user, and visualizations of models should help explain the rankings.
    \item \emph{G4: Determine the most preferable model, beyond the initial rankings:} In many cases, ranking is not enough to make judgment of the superior model. For example, in a classification problem of cancer related data, two models may have the same ranking based on the given accuracy criteria. However, the model with fewer false negative predictions might be preferable. Visualizations can provide an efficient way to communicate the capabilities of different models; even simple visualizations like colored confusion matrices offer much more information than a static metric score.
    \item \emph{G5: Compare model predictions on individual data points in the context of the input dataset:} Information about the model's predictions, such as their accuracies or their error, were difficult to extrapolate on without the context of individual data instances they predicted upon.  Users suggested that having the data overview and the model results on separate tabs of the system made this difficult.  Users want to judge model predictions in coordination with exploratory data analysis views.  Model explanation techniques such as those linking confusion matrix cells to individual data instances offer a good example of tight linking between the data space and the model space\cite{zhang2018manifold,amershi2015modeltracker,ren2017squares}.
    \item \emph{G6: Transition seamlessly from one step to another in the overall workflow:} Providing a seamless workflow in the resulting interface helps the user to perform the different tasks required in generating and selecting the relevant models. The system should guide the user in the current task as well as to transition it to the next task without any extra effort. Furthermore, useful default values (such as highly relevant problem specifications or the top ranked predictive model) should be provided for non-expert users so that they can finish at least the default steps in the EMA workflow.  Accompanying visualizations that dynamically appear based on the current workflow step can provide easy-to-interpret snapshots of what the system is doing at each step. 
\end{itemize}


\noindent It should be noted that our distilled set of requirements does not include participants' comments relating to data cleaning, data augmentation, or post-hoc manual parameters tuning of the selected models. While they are important to the users and relevant to their data analysis needs, these topics are familiar problems in visual analytics systems and are therefore omitted from consideration.

\subsection{Workflow Design}
\label{subsec:workflowDesign}


Based on the six identified task requirements, we propose a workflow as shown in Figure~\ref{fig:workflow}. The workflow consists of seven steps that are then grouped into three high-level tasks: data and problem exploration, model generation, and model exploration and selection. Below we detail each step of the workflow.





    \noindent \textbf{Step 1 -- Data Exploration:} 
    In response to \textbf{G1}, we identify data exploration as a required first step. Before a user can explore the model space, they must understand the characteristics of the data. Sufficient information needs to be presented so that the user can make an informed decision as to which types of predictions are suitable for the data. Furthermore, the user needs to be able to identify relevant attributes or subsets of data that should be included (or avoided) in the subsequent modeling process.


    
    \noindent \textbf{Step 2 -- Problem Exploration:} In response to \textbf{G1} and \textbf{G2}, we also identify the need of generating automatically a valid set of problem specifications. These problem specifications give the user an idea of the space of potential models, and they can use their understanding of the data from Step 1 to choose which are most relevant to them. 

    \noindent \textbf{Step 3 -- Problem Specification Generation: } In response to \textbf{G2} and \textbf{G3}, we identify the need of  generating a valid, machine-readable final set of problem specifications after the user explores the dataset and the automated generated problem specifications set. A EMA visual analytic system needs to provide the option to user to refine and select a problem specification from the system generated set or to add a new problem specification. Furthermore, the user should also be able to provide or edit performance metrics (such as accuracy, F1-score, mean squared root, etc.) for each problem specification. 


    \noindent \textbf{Step 4 -- Model Training and Generation:} The generated problem specifications will be used to generate a set of trained models. Ideally, the resulting set of models should be diverse. For example, for a classification problem, models should be generated using a variety of classification techniques (e.g. SVM, random forest, k-means, etc.). Since these techniques have different properties and characteristics, casting a wide net will allow the user to better explore the space of possible predictive models in the subsequent EMA process.

    \noindent \textbf{Step 5 -- Model Exploration:} In response to \textbf{G3}, we identify the need of presenting the resulting predictive models in some ranked form (e.g., based on either used performance metric or the time required in generating the model). An EMA visual analytics system needs to present the resulting models through some visualizations, e.g., a confusion matrix for a classification problem type or a residual bar chart for regression problem type (see Fig.~\ref{fig:workflow}(5)), so that the user can explore predictions of the models and facillitate comparisons between them. We also identify from \textbf{G5} that cross-linking between individual data points in a model and data exploration visualization would be useful for the user to better understand the model. It should be noted that a prerequisite for model exploration is to present models in an interpretable encoding, and the available encoding depends on the types of models being explored.  Lipton posited that there are two types of model interpretability: transparency, in which a model can be structurally interpreted, and post-hoc interpretability, in which a model is interpreted via its predictions on held out data~\cite{lipton2016mythos}.  In our workflow, because we aim to allow for any type of model, it is difficult to compare wildly different parts of the model space (a kNN model vs. a deep learning model) based on their structure.  Instead, we favor a post-hoc approach, where the models are explored via their predictions.  
    
 
    \noindent \textbf{Step 6 -- Model Selection:} In response to \textbf{G4} and \textbf{G5}, we identify the need for selecting the user's preferred models based on the model and data exploration. An EMA visual analytics system needs to provide the option to the user to select one or more preferable models in order to export for later usage. 

    \noindent \textbf{Step 7 -- Export Models:} In response to \textbf{G4}, we also identify that the user also requires to export the selected preferable models so that they can use them for future predictions.


    Finally, we identify from the response of \textbf{G6} that an EMA visual analytic system needs to make sure that the transition from one workflow step to another one should be seamless. We assume that any implementation of our proposed EMA workflow in Figure~\ref{fig:workflow} should supply such smooth transitions so that a non-expert user would also be able to finish the workflow from start to end.

\section{Iterative System Design and Evaluation}

\begin{table*}[htb]
\begin{center}
\begin{tabular}{l}
\centering
 \includegraphics[width=.95\linewidth]{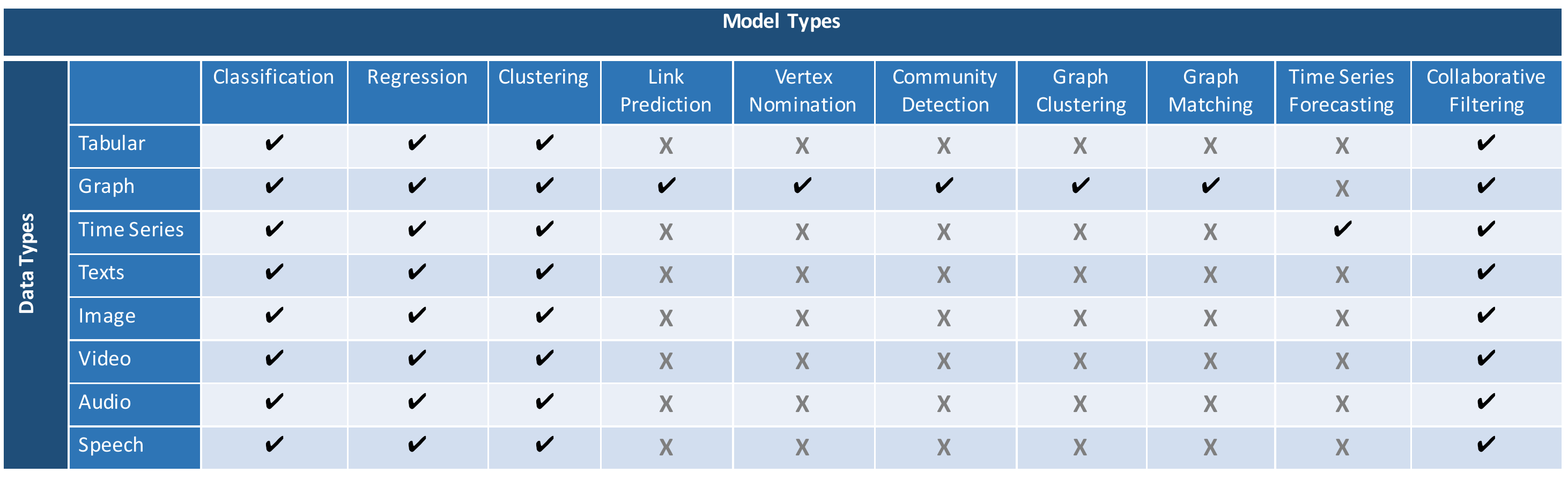}
 \end{tabular}
 \caption{
 List of all model types and data types supported by our experimental system. A check mark indicates if a model type can be applied to a particular data type, while a cross mark is used to denote incompatible matching between data and model types.}
 \label{tab:TaskAndData}
\end{center}
\end{table*} 

To validate our visual analytics workflow for EMA, we performed two rounds of iterative design and evaluation of the initial prototype system.  First, we describe the updated system used in the evaluation.
Due to confidentiality concerns, the screenshots shown in this paper use a set of publicly available datasets\footnote{https://gitlab.com/datadrivendiscovery/tests-data} that are different from the data examined by the subject matter experts during the two rounds of evaluations.

\subsection{Redesigned EMA System}
\label{subsec:experimentalSystem}


Our redesigned system used for the study significantly differs from the original prototype in three important ways. First, it fully supports the workflow as described in the previous section, including Problem Exploration and Problem Specification Generation. Second, the new system supports 10 types of models (compared to the prototype that only supported two). Lastly, in order to accommodate the diverse subject matter experts' needs, our system was expanded to support a range of input data types. Table~\ref{tab:TaskAndData} lists all of the supported model and data types of our redesigned system.

From a visual interface design standpoint, the new system also appears differently from the prototype.
The key reason for the interface redesign is to provide better guidance to users during the EMA process, and to support larger number of data and model types.
We realized during the redesign process that the UI of the original prototype (which used tabs for different steps of the analysis) would not scale to meet the requirements of addressing the seven steps of the EMA workflow.

\begin{figure*}[tb]
  \centering 
  \includegraphics[width=0.95\linewidth]{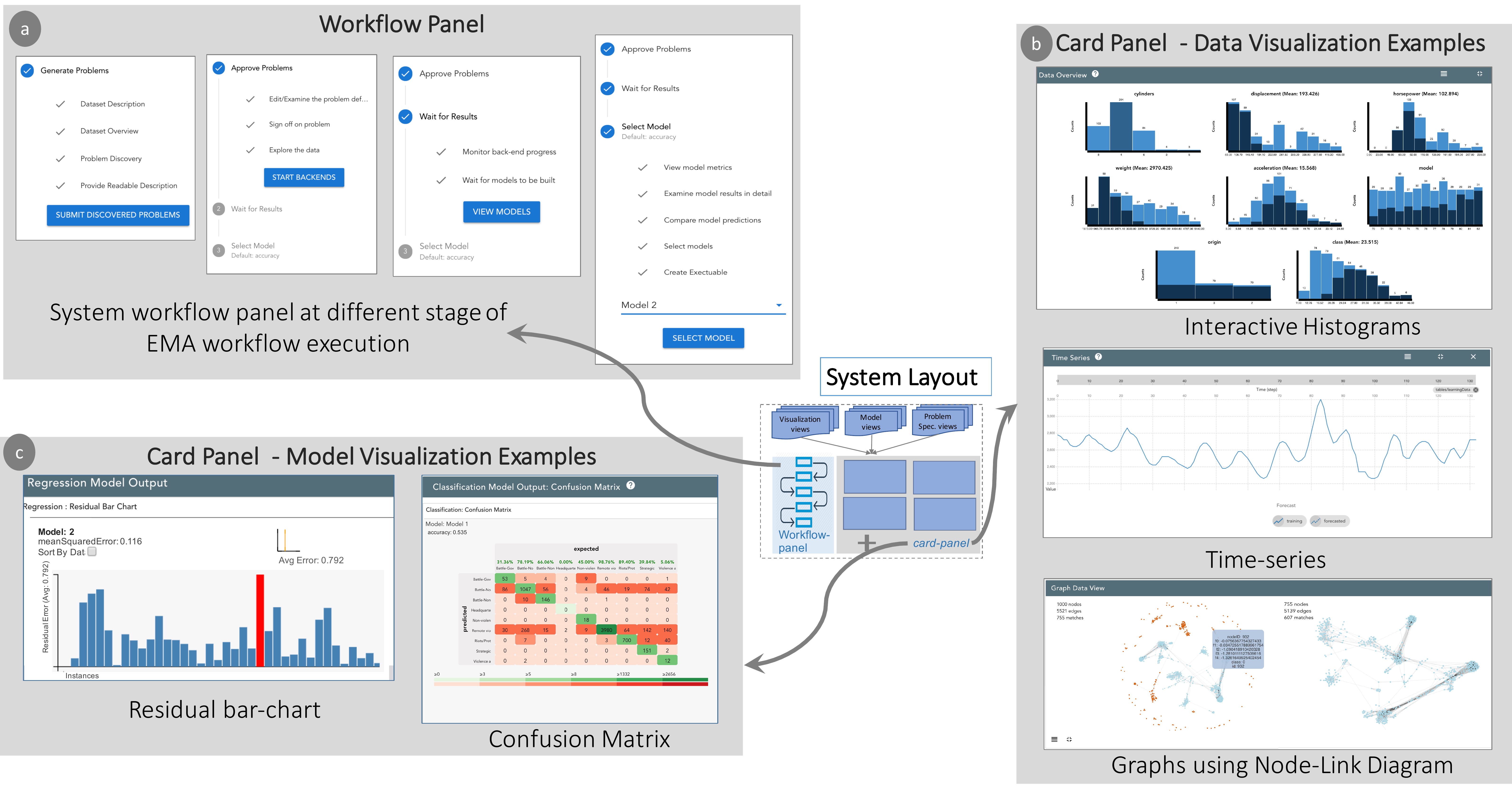}
  \caption{Components of the experimental system.  The box in the center shows the system layout, which consists of two parts, the left-side workflow panel and the right-side card panel. (a) shows EMA workflow at different stages in the experimental system, (b) shows three examples of data visualization cards, and (c) shows two examples of model visualization cards.}
  \label{fig:sysOverview}
\end{figure*}

Figure~\ref{fig:sysOverview} shows screenshots of components of the system that highlight the system's support for guiding the user through the steps of the EMA workflow.
The visual interface consists of two parts (see Fig.~\ref{fig:sysOverview}, where we provide the overall system layout in the center). The  
\emph{workflow-panel} (see Fig.~\ref{fig:sysOverview}(a)), positioned on the left side of the system layout, shows the current level and status of workflow execution. 
On the right side of the system layout, the \emph{card-panel} consists of multiple \emph{cards} where each card targets a particular step described in the EMA workflow.

\noindent \textit{Visualization Support for Data Exploration:}

\noindent For step one of the workflow, \textbf{data exploration}, the system renders several cards providing an overview of the dataset. This includes both a dataset summary card containing any metadata available, such as dataset description and source, as well as cards with interactive visualizations for each data type in the dataset (a few examples are provided in  Fig.~\ref{fig:workflow}(a) and in Fig.~\ref{fig:sysOverview}(b)). 
Currently, the system supports eight input data types: tabular, graph, time-series, text, image, video, audio, and speech.  Datasets are taken in as CSVs containing tabular data that can point to audio or image files, and rows can contain references to other rows, signifying graph linkages.  Data types (e.g., numeric, categorical, temporal, or external references) are explicitly provided - the system does no inference of data types. If a dataset contains multiple types of data, the system a specifically designed card for each type of data. In all cases, the user is also provided a searchable, sortable table showing the raw tabular data. All data views are cross-linked to facilitate insight generation. To limit the scope of the experimental system, our system is not responsible for data cleaning or wrangling, and it assumes that these steps have already been done before the system gets the data.

For example, in the case of only tabular data a set of cross-linked histograms are provided (see Fig.~\ref{fig:sysOverview}(b)), empowering the user to explore relationships between features and determine which features are most predictive. Furthermore, a searchable table with raw data fields is also provided. For graph data, node-link diagrams are provided (see Fig.~\ref{fig:sysOverview}(b)). Temporal data is displayed through one or more time-series line charts (see Fig.~\ref{fig:sysOverview}(b)), according to the number of input time-series. For textual data, the system shows a simple searchable collection of documents to allow the user to search for key terms. Image data is displayed in lists sorted by their labels. Audio and speech files are displayed in a grid-list format with amplitude plots, and each file can also be played in the browser through the interface. Video files are also displayed in a grid-list format, and can be played in the browser through the interface as well. In the case of an input dataset with multiple types of data, such as a social media networks where each row in the table references a single node in a graph, visualizations are provided for both types of data (e.g., histograms for table and node-link diagrams for graphs) and are cross-linked via the interaction mechanisms (i.e., brushing and linking). The exact choices for visual encodings for input dataset types are not contributions of this paper, and so mostly standard visualizations and encodings were used. 

\noindent \textit{Problem Specification Generation and Exploration:}

\noindent After data exploration, the user is presented with a list of possible problem specifications depending on the input dataset (step 2, \textbf{problem exploration} in the EMA workflow). This set is auto-generated by first choosing each variable in the dataset as the target variable to be predicted, and then generating a problem specification for each machine learning model type that is valid for that target variable. For example, for a categorical target variable, a classification problem specification is generated. For a numeric target variable, specifications are generated for both regression and collaborative filtering. Table~\ref{tab:TaskAndData} shows the relationships between an input dataset and the possible corresponding model types supported in our system. The system also generates different problem specifications for each metric valid for the problem type and the type of predicted variable (e.g., accuracy, f1 score, precision). Together, the target prediction variable, the corresponding model type, metrics, and features to be used for predicting the target prediction variable make up a \textit{problem specification}.

The user can select interesting problem specifications from the system-generated list of recommendations, and refine them by removing features as predictors.
Users can also decide to generate their own problem descriptions from scratch, in case non of the system-generated suggestions fit the their goals.
In either case, the next step of the EMA workflow is for the user to finalize the \textbf{problem specifications} (see Fig.~\ref{fig:workflow}(3)).
The resulting set of problem specifications is then used by backend autoML systems to generate the corresponding machine learning models.


\newpage
\noindent \textit{Visualization Support for Model Exploration and Selection:}

\noindent Our system's support for \textbf{model generation} (step 4 of the EMA workflow) relies on the use of an automated machine learning (autoML) library, developed under the DARPA D3M program~\cite{shen_d3m}. These autoML systems are accessible through an open source API\footnote{https://gitlab.com/datadrivendiscovery/ta3ta2-api} based on the gRPC protocol\footnote{https://grpc.io/}. 
An autoML system requires the user to provide a well-defined problem specification (i.e., the target prediction variable, the model type, the list of features to be used for training the model, the performance metrics) and a set of training data. It then automatically searches through a collection of ML algorithms and their respective hyperparameters, returning the ``best'' models that fit the user's given problem specification and data. Different autoML libraries such as AutoWeka\cite{thornton2013auto, kotthoff2016auto}, Hyperopt\cite{bergstra2013hyperopt, komer2014hyperopt}, and Google Cloud AutoML\cite{googlecloudautoml} are in use either commercially or as open source tools. 
Our system is designed to be compatible with several autoML libraries under the D3M program, including\cite{jin2018efficient, sheni2018prediction}. Note that the sampling of models is entirely driven by the connected autoML systems, and our system does not encode any instructions to the autoML systems beyond the problem specification chosen by the user. However, the backends we connect to generate diverse, complex models, and automatically construct machine learning pipelines including feature extraction and dimensionality reduction steps.  

Given the set of problem specifications identified by the user in the previous step, the autoML library automatically generates a list of candidate models.
The candidate models are then visualized in an appropriate interpretable representation of their predictions, corresponding to the modeling problem currently being explored by the user (step 5, \textbf{model exploration}). All types of classification models, including multiclass, binary, and variants on other types of data such as community detection, are displayed to the user as interactive confusion matrices (see Fig.~\ref{fig:sysOverview}(c)). Regression models and collaborative filtering models are displayed using sortable interactive bar charts displaying residuals (see Fig.~\ref{fig:sysOverview}(c)). Time-series forecasting models are displayed using line charts with dotted lines for predicted points. Cross-linking has been provided between individual data points on these model visualizations and the corresponding attributes in the data exploration visualizations of the input dataset. Furthermore, cross-linking between the models has also been provided to help the user in comparing between the generated models.  

Our system initially shows only the highest ranked models produced by the autoML library, as the generated models could be in the hundreds in some cases. This ranking of models is based on the user selected metric in the problem specification. 

After a set of suggested models had been generated and returned by the autoML engine, the system provides views to inspect the model's predictions on holdout data. Using this information, they select one or more specific models and request the autoML library to export the selected model(s) (Steps 6 and 7, \textbf{model selection} and \textbf{export models}).



\subsection{Evaluation}

To evaluate the validity of our proposed EMA workflow and the efficacy of the prototype system, we conducted two rounds of evaluations.  Similar to the feedback study, the participants of these two rounds of evaluation were also recruited by NIST. 
Five subject matter experts participated in the first round of evaluation, and four participated in the second.  One participant in the second round was unable to complete the task due to connectivity issues.
None of the experts participated in both studies (and none of them participated in the previous feedback study).  The two groups each used different datasets, in an aim to test out the workflow in differing scenarios.    

\noindent \textbf{Method:} Several days prior to each experiment, participants were part of a teleconference in which the functioning of the system was demonstrated on a different dataset than would be used in their evaluation. They were also provided a short training video\cite{snowcat_training_video} and a user manual\cite{snowcat_training_manual} describing the workflow and individual components of the system they used.  

For the evaluation, participants were provided with a link to a web interface through which they would do their EMA. They were asked to complete their tasks without asking for help, but were able to consult the training materials at any point in the process. The modeling specifications discovered by users were recorded, as well as any exported models. After completing their tasks, participants were given an open-ended questionnaire about their experience.  After the first round of evaluation, some user feedback was incorporated into the experimental system, and the same experiment was held with different users and a different dataset.  All changes made to the experimental system were to solve usability issues, in order to more cleanly enable users to follow the workflow presented in this work.

\noindent \textbf{Tasks:} In both evaluation studies, participants were provided with a dataset on which they were a subject matter expert. They were given two successive tasks to accomplish within a 24-hour period. The first task was to explore the given dataset and come up with a set of modeling specifications that interested them.  The second task supplied them with a problem specification, and asked them to produce a set of candidate models using our system, explore the candidate models and their predictions, and finally choose one or more models to export with their preference ranking. Their ranking was based on which models they believed would perform the best on held out test data.  The two tasks taken together encompass the workflow proposed in this work.  
The problem specifications discovered by participants were recorded, as well as the resulting models with rankings exported by the participants.

\subsection{Findings}
All 8 of the participants were able to develop valid modeling problems and export valid predictive models.  Participants provided answers to a survey asking for their top takeaways from using the system to complete their tasks.  They were also asked if there were additional features that were missing from the workflow of the system.  We report common comments on the workflow, eliding comments pertaining to the specific visual encodings used in the system.

\noindent \textbf{Efficacy of workflow:}  Participants felt that the workflow was successful in allowing them to generate models. One participant noted that the workflow ``\textit{...can create multiple models quickly if all (or most data set features are included... [the] overall process of generating to selecting model is generally easy}''.  Another participant agreed, stating that ``\textit{The default workflow containing comparisons of multiple models felt like a good conceptual structure to work in}.''

The value of individual stages of the workflow were seen as well: ``\textit{The problem discovery phase is well laid out.  I can see all the datasets and can quickly scroll through the data}''.  During this phase, participants appreciated the ability to use the various visualizations in concert with tabular exploration of the data, with one participant stating that ``\textit{crosslinking visualizations [between data and model predictions] was a good concept}'', and another commenting that the crosslinked tables and visualizations made it ``\textit{very easy to remove features, and also simple to select the problem}.''

\noindent \textbf{Suggestions for implementations:}  Participants were asked what features they thought were most important for completing the task using our workflow.  We highlight these so as to provide guidance on the most effective ways to implement our workflow, and also to highlight interesting research questions that grow out of tools supporting EMA.  

Our experimental system allowed for participants to select which features to use as predictor features (or independent variables) when specifying modeling problems.  This led several participants to desire more sophisticated capabilities for feature generation, to ``\textit{create new derivative fields to use as features}''.  

One participant noted that some of the choices in generating problem specifications were difficult to make without first seeing the resulting models, such as the loss function chosen to optimize the model. The participant suggested that, rather than asking the user to provide whether root mean square error or absolute error is used for a regression task, that the workflow ``\textit{have the system combinatorically build models for evaluation (for example, try out all combinations of ``metric'')}''.  This suggests that the workflow can be augmented by further automating some tasks.  For example, some models could be trained before the user becomes involved, to give the user some idea of where to start in their modeling process.   

The end goal of EMA is one or more exported models, and several participants noted that documentation of the exported models is often required.  One participant suggested the system could ``\textit{export the data to include the visualizations in reports}''.  This suggests that an implementation of our workflow should consider which aspects of provenance it would be feasible to implement, such as those expounded on in work by Ragan et al.\cite{ragan2016characterizing}, in order to meet the needs of data modelers.  
Another participant noted that further ``\textit{understanding of model flaws}'' was a requirement, not only for the sake of provenance, but also to aid in the actual model selection. 
Model understandability is an open topic of research\cite{gleicher2013explainers}, and instance-level tools such as those by Ribeiro et al.\cite{ribeiro2016should} and Krause et al.\cite{krause2017workflow} would seem to be steps in the right direction.  
Lastly, it was noted that information about how the data was split into training and testing is very relevant to the modeler.  Exposing the training/testing split could be critical if there is some property in the data that makes the split important (i.e. if there are seasonal effects in the data).  

\noindent \textbf{Limitations of the Workflow:} The participants noted that there were some dangers in developing a visual analytics system that enabled exploratory modeling, noting that ``\textit{simplified pipelines like those in the task could very easily lead to serious bias or error by an unwary user (e.g. putting together a causally nonsensical model)}''.  The ethics of building tools that can introduce an untrained audience to new technology is out of the scope of this work, but we do feel the topic is particularly salient in EMA, as the resulting models will likely get deployed in production.  
We also contend that visual tools, like those supported by our workflow, are preferable to non-visual tools in that the lay user can get a sense of the behavior of models and the training data visually.  It could be that additional safeguards should be worked into the workflow to offer a sort of spell-check of the models, similar to how Kindlmann and Scheidegger recommend that visualizations are run through a suite of sanity checks before they are deployed\cite{kindlmann2014algebraic}.  

The same participant also noted that streamlining can also limit the ability of the user if they are skilled: ``\textit{it doesn't provide sufficient control or introspection... I wanted to add features, customize model structure, etc., but I felt prisoner to a fixed menu of options, as if I was just a spectator}''.  
While some of this can be ameliorated by building a system more angled at the expert user and including more customization options, ultimately the desired capabilities of a system by an expert user may be beyond the ceiling of the workflow we have presented.

\section{Usage Scenarios}
In this section, we offer two examples of how the system might be used for exploratory model analysis.  
Through these two scenarios we explain the role of the user during each step of the workflow.
The first scenario involves the exploration of a sociological dataset of children's perceptions of popularity and the importance of various aspects of their lives.  It is used to build predictive models which can then be incorporated into an e-learning tool.
The second scenario requires building predictive models of automobile performance for use in prototyping and cost planning.

\label{sec:usecase}
\subsection{Analyzing the Popular Kids Dataset}

The \textit{Popular Kids}\footnote{http://tunedit.org/repo/DASL} dataset consists of 478 questionnaires of students in grades 4, 5, and 6 about how they perceive importance of various aspects of school in the popularity of their peers. The original study found that among boy respondents, athletic prowess is perceived as most important for popularity, while among girl respondents, appearance is perceived as most important for popularity~\cite{chase1992role}.

\textit{John} works for a large public school district that is trying to determine what data to collect for students on an e-learning platform.  Project stakeholders believe that they have some ability to gather data from students in order to personalize their learning plan, but that gathering too much data could lead to disengagement from students.  Therefore, John must find what sorts of predictive models can be effective on data that is easy to gather.  

John downloads the Popular Kids dataset and loads it into the application.  The system shows him linked histograms of the various features of the dataset, as well as a searchable table.  He explores the data (\textbf{Step 1 in EMA workflow}),  noting that prediction of a student's belief in the importance of grades would be a valuable prediction for the e-learning platform.  He scans the list of generated problems (\textbf{Step 2}), selecting a regression problem predicting the belief in grades.  He refines the problem (\textbf{Step 3}, removing variables in the training set of which school the student was from, since that data would not be relevant in his deployment scenario.  He then sends this problem to the autoML backend, which returns a set of models (\textbf{Step 4}).  The system returns to him a set of regression models (\textbf{Step 5}), displayed as bar charts showing residuals on held out data (see Figure~\ref{fig:regression}.  He notes that none of the regression models have particularly good performance, and in particular, by using cross linking between the regression models and the raw data visualizations, he notes that the resulting models have much more error on girls than on boys.  

\begin{figure}[tb]
  \centering 
  \includegraphics[width=1\linewidth]{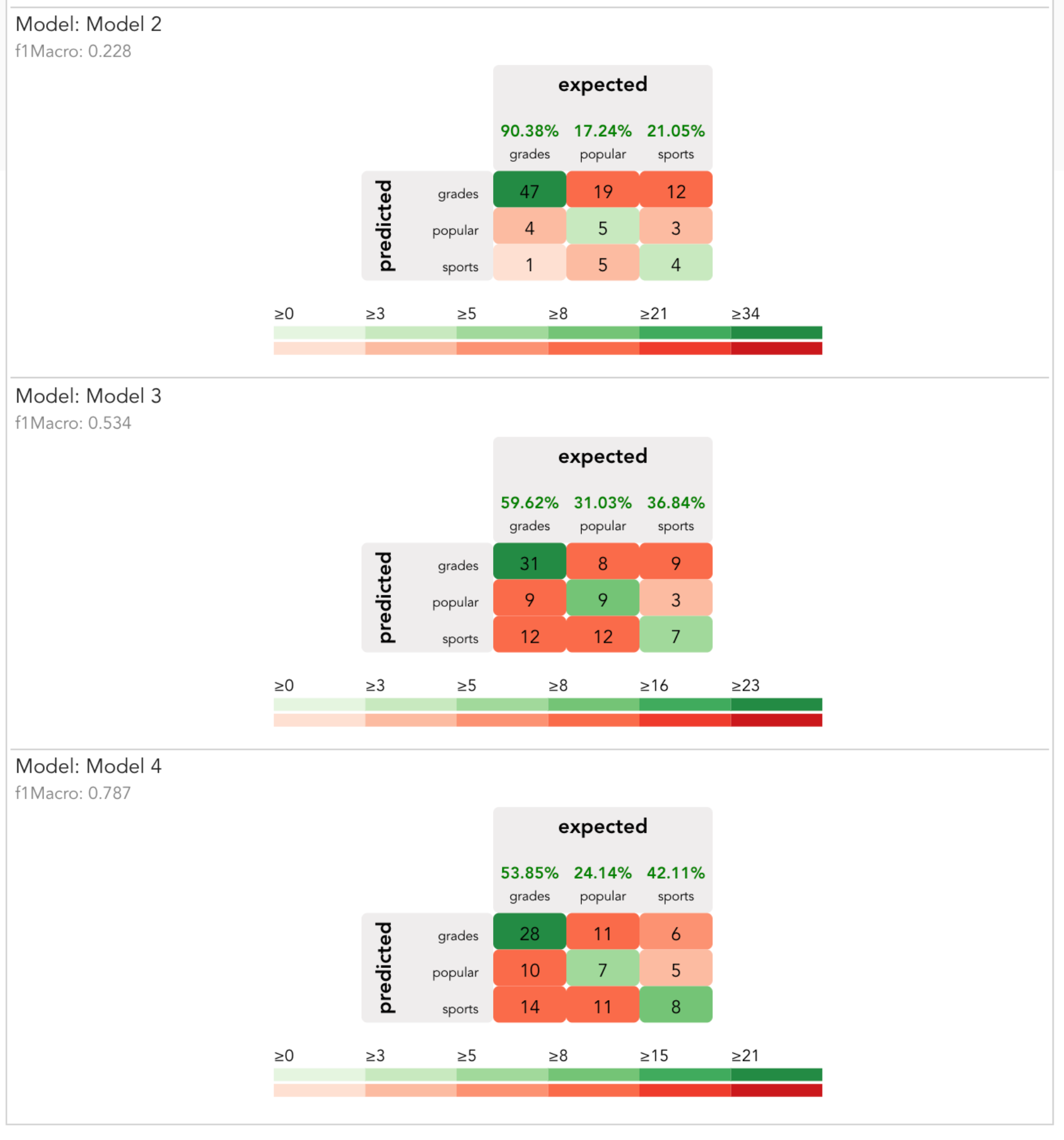}
  \caption{A set of confusion matrices showing the different classification models predicting \textit{Goal} of students in the \textit{Popular Kids} dataset~\cite{chase1992role}.  The user is able to see visually that, while the middle model has the highest overall accuracy, it performs much better on students who have high grades as their goal.  Instead, the user chooses the bottom model, because it performs more equally on all classes.
}
  \label{fig:kids_example}
\end{figure}

At this point, John determines that the dataset is not particularly predictive of belief in grades, and decides to search for another predictive modeling problem.  He returns to \textbf{Step 3} and scans the set of possible problems.  He notes that the dataset contains a categorical variable representing the student's goals, with each student marking either \textit{Sports}, \textit{Popular}, or \textit{Grades} as their goal.  He chooses a classification problem, predicting student goal, and removes the same variables as before.  He submits this new problem and the backend returns a set of models (\textbf{Step 4}).  The resulting classification models are visualized with a colored confusion matrix, seen in figure~\ref{fig:kids_example}.  John compares the different confusion matrices (\textbf{Step 5}), and notes that even though model 2 is the best performing model, it performs poorly on two out of the three classes.  Instead, he chooses model 3, which performs farily well on all three classes (\textbf{Step 6}).  He exports the model (\textbf{Step 7}), and is able to use it on data gathered by the e-learning platform.

\subsection{Modeling Automobile Fuel Efficiency}
\emph{Erica} is a data scientist at an automobile company and she would like to develop predictive models that might anticipate the performance or behavior of a car based on potential configurations of independent variables. 
In particular, she wants to be able to predict how various designs and prototypes of vehicles might affect properties of the car that affect its sales and cost.
She hopes to discover a model that can be used to assess new designs and prototypes of vehicles, before they are built. 

Erica has access to a dataset containing information about 398 cars (available from OpenML\cite{OpenML2013}), and she would like to build a set of predictive models using different sets of prediction features to determine which features may be most effective at predicting fuel efficiency. She begins by loading the \textbf{Data Source} and explores the relationship between attributes in the histogram view (\textbf{Step 1}), shown in the top histogram visualization in Figure~\ref{fig:sysOverview}(b).
By hovering over the bars corresponding to mpg, she determines that the number of cylinders and the class may be good predictors. She then explores the system generated set of problem specifications (\textbf{Step 2}).
She looked on all the generated problem specifications with ``class'' as predicting feature. She decides on predicting miles per gallon, and selects a regression task. She selects the provided default values for the rest of the problem specification (\textbf{Step 3}). 

\begin{figure}[tb]
  \centering 
  \includegraphics[width=0.9\linewidth]{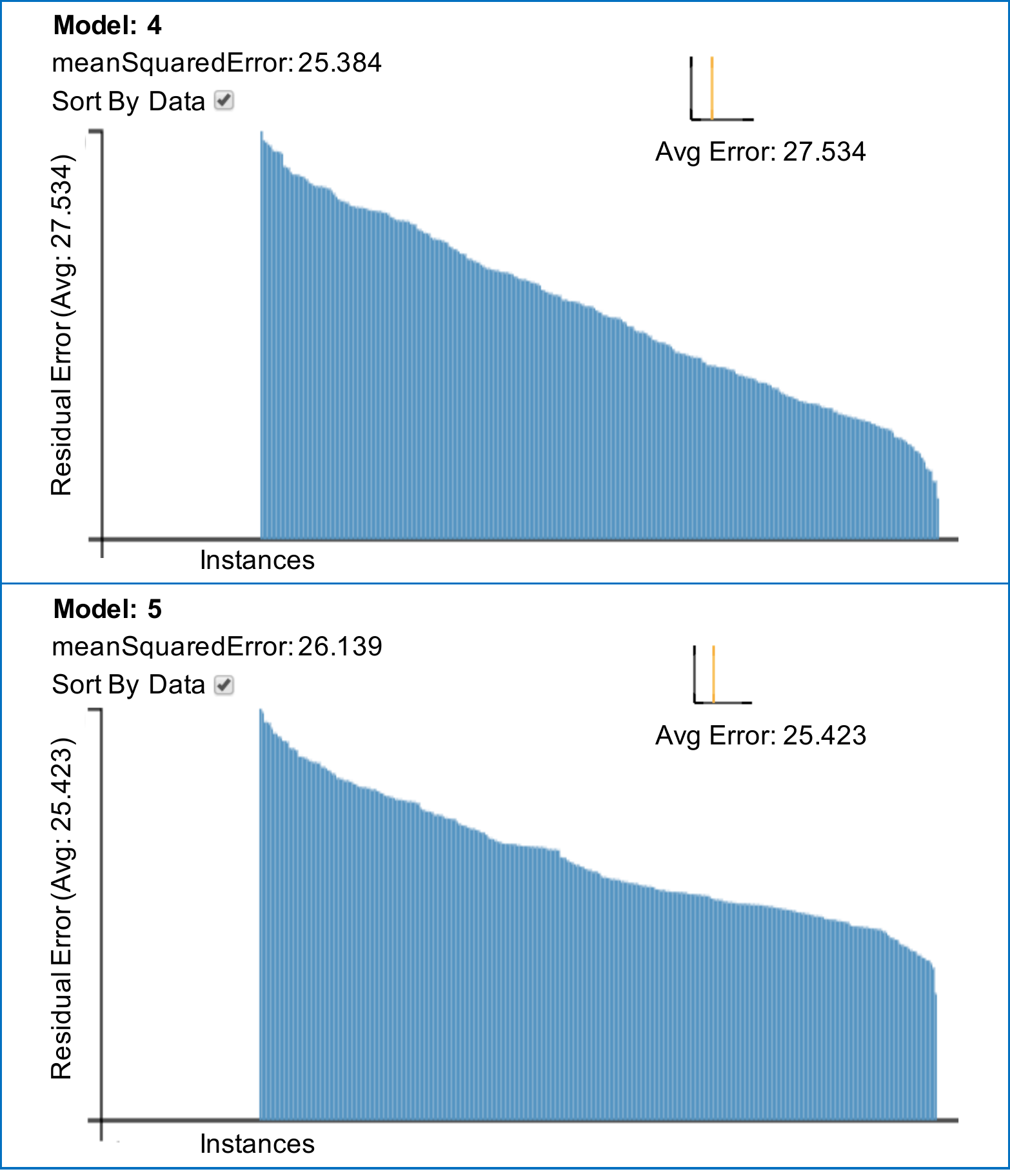}
  \caption{Regression plots of the two best models returned by a machine learning backend.}
  \label{fig:regression}
\end{figure}

The ML backend trains on the given dataset and generates six models (\textbf{Step 4}). Erica starts to explore the generated regression models, visualized through residual bar charts (\textbf{Step 5}). The model visualization in this case gives Erica a sense of how the different models apportion residuals by displaying a bar chart of residuals by instance, sorted by the magnitude of residual (see Fig.~\ref{fig:regression}).   

Erica notices that the two best models both have similar scores for mean squared error.
She views the residual plots for the two best models, and notes that, while the mean squared error of model 4 is lowest, model 5 apportions residuals more evenly among its instances (see Fig.~\ref{fig:regression}).
Based on her requirements, it is more important to have a model that gives consistently close predictions, rather than a model that performs well for some examples and poorly for others. Therefore, she selects the model 5 (\textbf{Step 6}) to be exported by the system (\textbf{Step 7}).
By following the proposed EMA workflow, Erica was able to get a better sense of her data, to define a problem, to generate a set of models, and to select the model that she believed would perform best for her task.

\section{Conclusion}
In this work, we define the process of exploratory model analysis (EMA), and contribute a visual analytics workflow that supports EMA.  We define EMA as the process of discovering and selecting relevant models that can be used to make predictions on a data source.  In contrast to many visual analytics tools in the literature, a tool supporting EMA must support problem exploration, problem specification, and model selection in sequence.  Our workflow was derived from feedback from a pilot study where participants discovered models on both classification and regression tasks.

To validate our workflow, we built a prototype system and ran user studies where participants were tasked with exploring models on different datasets.  Participants found that the steps of the workflow were clear and supported their ability to discover and export complex models on their dataset.  Participants also noted distinct manners in which how visual analytics would be of value in implementations of the workflow.  We also present two use cases across two disparate modeling scenarios to demonstrate the steps of the workflow.
By presenting a workflow and validating its efficacy, this work lays the groundwork for visual analytics for exploratory model analysis through visual analytics.

\section{Acknowledgements}

We thank our collaborators in DARPA's Data Driven Discovery of Models (d3m) program.  This work was supported by National Science Foundation grants IIS-1452977, 1162037, and 1841349, as well as DARPA grant FA8750-17-2-0107.


\bibliographystyle{eg-alpha-doi}

\bibliography{ref}

\end{document}